# Mathematical modeling of evolution of horizontally transferred genes

Artem S. Novozhilov, Georgy P. Karev, and Eugene V. Koonin*

National Center for Biotechnology Information, National Library of Medicine, National Institutes of Health, Bethesda, MD 20894

*To whom correspondence should be addressed. Email koonin@ncbi.nlm.nih.gov

**Abstract**

We describe a stochastic birth-and-death model of evolution of horizontally transferred genes in microbial populations. The model is a generalization of the stochastic model described by Berg and Kurland and includes five parameters: the rate of mutational inactivation, selection coefficient, invasion rate (i.e., rate of arrival of a novel sequence from outside of the recipient population), within-population horizontal transmission ("infection") rate, and population size. The model of Berg and Kurland included four parameters, namely, mutational inactivation, selection coefficient, population size, and "infection". However, the effect of "infection" was disregarded in the interpretation of the results, and the overall conclusion was that horizontally acquired sequences can be fixed in a population only when they confer a substantial selective advantage onto the recipient and therefore are subject to strong positive selection. Analysis of the present model in different domains of parameter values shows that, as long as the rate of within-population horizontal transmission is comparable to the mutational inactivation rate and there is even a low rate of invasion, horizontally acquired sequences can be fixed in the population or at least persist for a long time in a substantial fraction of individual in the population even when they are neutral or slightly deleterious. The available biological data strongly suggest that intense within-population and even between-populations gene flows are realistic for at least some prokaryotic species and environments. Therefore our modeling results are compatible with the notion of a pivotal role of horizontal gene transfer in the evolution of prokaryotes.

**Introduction**

Sequencing of multiple, complete genomes of diverse life forms and the ensuing advent of comparative genomics have dramatically changed the prevailing picture of evolution, at least for the prokaryotic world. It became clear that the evolutionary process is much more flexible and dynamic than previously imagined. In addition to the vertical inheritance of genes along a tree-like evolutionary trajectory, lineage-specific gene loss and horizontal (lateral) gene transfer (HGT) have emerged as major evolutionary forces, leading to the ideas of "uprooting the tree of life" and the concept of "horizontal genomics" (Pennisi 1998; Doolittle 1999a; Doolittle 1999b; Pennisi 1999; Doolittle 2000; Koonin, Aravind, and Kondrashov 2000; Koonin, Makarova, and Aravind 2001). Under a somewhat extreme view of the prevalence of HGT in the evolution of prokaryotes, even coherent tree topologies observed in mutligene analyses might be due to gradients of HGT propensity permeating the prokaryotic world: stable clusters in such trees are thought to comprise groups of microbes which exchange genes frequently (Gogarten, Doolittle, and Lawrence 2002). However, the extent of HGT in prokaryotic evolution remains a matter of contention (Kurland 2000; Kurland, Canback, and Berg 2003). Like other events that happened in the evolutionary past, each individual case of HGT is hard to prove beyond reasonable doubt. Relatively recent cases of probable HGT are usually demonstrated through anomalous nucleotide composition or codon usage of the genes thought to have been transferred (Tsirigos and Rigoutsos 2005). These methods are not applicable to putative ancient transfers which are detected on the basis of unexpected phyletic patterns of genes (i.e., patterns of presence-absence in genomes from different taxa) and/or discrepancies in the topologies of phylogenetic trees (Ochman, Lawrence, and Groisman 2000; Koonin, Makarova, and Aravind 2001; Ragan 2001; Lawrence and Hendrickson 2003). In many studies, explicit phylogenetic analysis is complemented or replaced by simpler analyses of sequence similarity rankings, usually, based on BLAST search results (Koonin et al. 1997; Aravind et al. 1998). None of these approaches is free of substantial caveats (Kurland, Canback, and Berg 2003). Any phyletic pattern, however unusual, in principle, can be explained solely through multiple, lineage-specific gene losses (Koonin 2003). Similarly, phylogenetic tree topologies and sequence similarity rankings are strongly affected both by gene loss and by unequal rates of evolution in different lineages (Koski and Golding 2001). As a result, it has been posited that the notion that HGT dominates prokaryotic evolution is a misconception stemming from non-critical data analysis which suggests vastly exaggerated rates of HGT (Kurland 2000; Kurland, Canback, and Berg 2003).

As noticed in a recent review by Lawrence and Hendrickson, the study of HGT remains a research field in its adolescence (Lawrence and Hendrickson 2003). This does not seem surprising given that the appreciation of the potential major significance of HGT as an evolutionary factor dates only from the last three-four years of the 20$^{th}$ century when systematic comparison of multiple sequenced genomes became possible (Koonin et al. 1997; Doolittle 1999b; Koonin, Aravind, and Kondrashov 2000). The lack of certainty regarding the true extent of HGT is one of the crucial aspects of this immaturity of "horizontal genomics". The other distinct but not unrelated aspect is the need to integrate HGT into the framework of the existing evolutionary theory which, in its current form, is based solely on the notion of vertical inheritance of genetic characters (Kimura 1983). Specifically, it is necessary to identify the neutral and/or selective evolutionary factors that affect the fate of a horizontally transferred gene leading to its fixation in or elimination from a recipient population. It seems likely that a robust evolutionary theory of HGT will provide feedback for assessing evidence of individual HGT cases and ultimately for more reliable estimates of the role of this phenomenon in evolution. The first evolutionary-theoretical analysis of HGT in microbes has been reported by Berg and Kurland (Berg and Kurland 2002). They concluded that, at least in populations of large effective size that are typical of most prokaryotes, horizontally acquired genes can persist and get fixed only when they provide a strong selective advantage to the recipient. It seems that cases when horizontally acquired genes are strongly beneficial to the recipient should be exceedingly rare.

Indeed, when HGT occurs, a gene moves from one cellular background to another, especially when the donor and acceptor are taxonomically and biologically distant. Since cellular components are fine-tuned by natural selection for coordinate functioning, chances that an alien protein confers a selective advantage onto the recipient will be low. This would apply to different functional systems of the cell to a different degree, with those that are based on large, multiprotein complexes being affected most strongly as captured in the complexity hypothesis of Lake and coworkers (Jain, Rivera, and Lake 1999). Nevertheless, the biologically reasonable expectation seems to be that cases of unequivocally beneficial HGT should be (extremely) rare. Exceptions are known, e.g., situations when a horizontally acquired gene makes the recipient resistant to an antibiotic(s), allows it to occupy a new nutritional niche or makes it a pathogen(Brown, Zhang, and Hodgson 1998; Mazel and Davies 1999; Rowe-Magnus, Davies, and Mazel 2002). Nevertheless, combination of this biological reasoning and the theoretical conclusions of Berg and Kurland fuels the contention that the extent of HGT in prokaryotes might have been seriously over-estimated by the proponents of "horizontal genomics" and that HGT is, perhaps, an important but not a decisive factor of prokaryotic evolution (Kurland, Canback, and Berg 2003).

Here we develop more general theoretical models of HGT between microbial populations and identify the conditions under which fixation of neutral or even slightly deleterious horizontally transferred genes is possible.

## Results and discussion

### 1. Deterministic model

We consider a haploid asexual population with overlapping generations (continuous time). In this section, we assume that the effect of stochastic evolutionary factors, such as genetic drift, is negligible. It is assumed that the population under consideration consists of two types of individuals: those that carry a particular novel sequence and those that do not. This means that the state of a population with regard to the novel sequence in question at a given moment $t$ is completely determined by two numbers $n_1(t)$ and $n_2(t)$ (the numbers, respectively, of the first- and second-type individuals at the moment $t$) or just one number $p(t) = n_1(t)/(n_1(t) + n_2(t))$, where $p(t)$ is the fraction of the first-type individuals in the population.

The assumptions of the simplest model are as follows. A novel sequence can be transmitted vertically through cell division or acquired via HGT. The novel sequence is also subject to mutational inactivation (point mutations, insertions and deletions) with a constant rate $u$. We assume that Malthusian parameters (intrinsic growth rates) are $m_1$ and $m_2$ (Nagylaki 1992). In addition, we assume that there is invasion of the first-type individuals with the rate $\gamma N$ where $N$ is the total size of the population, $N = n_1 + n_2$. Specifically, with regard to HGT, invasion means that acquisition of a particular alien gene is not a unique event, with a negligibly small probability of repeated occurrence, but rather that there is continuous influx of the alien gene, even if occurring at a low rate.

Taking into account only processes of mutation, selection, and invasion, the textbook system (Nagylaki 1992) is obtained for the numbers of individuals of different types:

$$\dot{n}_1 = m_1 n_1 - u n_1 + \gamma N,$$
$$\dot{n}_2 = m_2 n_2 + u n_1,$$

or for frequency $p$:

$$\dot{p} = (m_1 - \overline{m}) p - u p + \gamma q,$$

where $q = 1 - p$ and $\overline{m}$ is the mean continuous fitness, $\overline{m} = m_1 p + m_2 q$. This is the simplest mathematical model describing changes in the genetic content of a population. Note that, in this model, invasion is a special case of recurrent mutation with the rate $\gamma$.

In terms of frequencies of different types of individuals in the population, the assumption that invasion occurs with the rate $\gamma N$ is equivalent to replacing the second-type individuals with first-type individuals (those that carry the novel sequence) with the rate $\gamma$. To be more precise, if we consider the following system of differential equations:

$$\dot{n}_1 = m_1 n_1 - u n_1 + \gamma n_2,$$
$$\dot{n}_2 = m_2 n_2 + u n_1 - \gamma n_2,$$

the equation for the frequency $p$ of first-type individuals remains unchanged.

There are three principal mechanisms for transfer of genes between microbes: (i) transduction, bacteriophage-mediated gene transfer, (ii) transformation, transfer of DNA (e.g., released from dead microbes) from the environment to a recipient cell, and (iii) conjugation, direct transfer of DNA from one cell to another mediated by a plasmid (Bushman 2001). Each of these processes can mediate HGT both within a microbial population (we will call this "infection" for short) and between populations (invasion). Generally, it is expected that infection rates are substantially higher than invasion rates.

We can define the infection rate as the probability per unit of time for an individual that does not have the novel sequence to acquire it (to become "infected"). As an analogue of the law of mass action in chemical kinetics, we assume that the infection rate is proportional to $n_1$, with the proportionality constant $\theta$. The intensity of contact between infected and uninfected organisms is proportional to $n_1 n_2 / N = n_1 n_2 /(n_1 + n_2)$. This directly applies to conjugation, which involves physical contact between two cells, but generally holds true for transduction and transformation as well inasmuch as the release of the novel sequence within a transdusing agent or transforming DNA is proportional to the number of cells which harbor that sequence. Combining all the above assumptions, we obtain the following deterministic HGT-selection-mutation-invasion model:

$$\dot{n}_1 = m_1 n_1 - u n_1 + \gamma N + \theta \frac{n_1 n_2}{n_1 + n_2},$$
$$\dot{n}_2 = m_2 n_2 + u n_1 - \theta \frac{n_1 n_2}{n_1 + n_2},$$
(1)

or

$$\dot{p} = (m_1 - \bar{m})p - up + \gamma q + \theta pq \qquad (2)$$

for the frequency of the infected individuals.

Assuming $m_1 - m_2 = s$, where $s$ is the selection coefficient of the infected individuals, we obtain a logistic-type equation with immigration:

$$\dot{p} = (s + \theta)p(1 - p) - up + \gamma(1 - p) \qquad (3)$$

that can be easily analyzed (e.g., (Matis and Kiffe 1999)). Note that the rate $\gamma$ is the rate at which uninfected individuals become infected which, as outlined above may involve different biological mechanisms. If all the parameters have non-zero values and $0 \leq p \leq 1$, Eq.(3) has only one equilibrium solution satisfying the condition $0 < p^* < 1$, and this equilibrium is asymptotically stable. Thus, under this model, a gene acquired via HGT will persist in the population as long as there is continuous (even if low-rate) invasion, i.e., influx of the novel sequence. However, even if $\gamma = 0$, i.e., there is no invasion and $s < 0$, i.e., the novel sequence is deleterious, but $\theta > -s$, there is a stable interior equilibrium that corresponds to the HGT-selection-mutation balance. This describes another scenario (distinct from the scenario with a selectively advantageous gene acquired) for persistence of an acquired gene even without continuous influx: the novel gene will persist if there is a sufficiently effective within-population mode of transmission (infection).

Equations (1)-(3) are classical models of population genetics of asexual organisms. The inclusion of the processes of infection and invasion and application of the law of mass action allows us to incorporate HGT within the bounds of known mathematical models.

**2. Stochastic model**

The simple, deterministic model described in the previous section includes only systematic factors of evolution whose rates are assumed to be constant (Wright 1949). To model genetic drift in the population, we formulate a stochastic counterpart of the model (1)-(3). To make the model mathematically manageable, we assume that the total size of the population is constant and equals $N$. We use a Markov birth and death process $\{X(t), t \geq 0\}$ with the finite state space $\{0,1,...,N\}$. The number, $n$, of individuals that carry a particular novel sequence (a gene acquired via HGT) determines the state of the population. Transitions in the process are only allowed to neighbor states. The rate of transition from state $n$ to state $n+1$ (the "birth rate" of infected individuals) is denoted $\lambda_n$, and the rate of transition from the state $n$ to the state $n-1$ is denoted by $\mu_n$ (the "death rate" of infected individuals, i.e., the rate at which the novel sequence is lost, which includes actual cell death among other processes).

The Kolmogorov forward equations for the state probabilities $p_n(t) = \Pr\{X(t) = n\}$ can be written as

$$\dot{p}_n = \mu_{n+1} p_{n+1} - (\lambda_n + \mu_n) p_n + \lambda_{n-1} p_{n-1}, \qquad n = 0,1,...,N.$$

In order to make sense of this equation, we put formally $\lambda_{-1} = \mu_{N+1} = p_{-1}(t) = p_{N+1}(t) = 0$. The state probabilities depend on the initial distribution $p_n(0)$.

This scheme corresponds to a Moran model (Moran 1958) which, in the case of a haploid population with overlapping generations, is more realistic for microbial populations than the more widely used Wright-Fisher model; in particular, a Moran model was adopted in the work of Berg and Kurland (Berg and Kurland 2002).

### *2.1 Statement of the model*

In order to analyze the model, we need to identify all the rates of the birth and death process. Let the individuals without the novel sequence divide with rate $\nu$; we will define the time unit in the model such that $\nu = 1$. Then, the individuals with the novel sequence divide with the rate $(1+s)$, where $s$ is the selection coefficient. When one of the individuals divides, we remove a random individual from the population to keep its size constant. As in the deterministic model (1)-(3), we take into account processes of inactivating mutation with the constant rate $u$, invasion with the constant rate $\gamma N$, and within-population transmission of the novel sequence (infection) with the rate $\theta$.

The system can change its state from $n$ to $n+1$ if an infected cell divides and there is no inactivating mutation or if a new infected cell immigrates. In this case, the total number of cells is $N+1$ and we need to remove one of the $N-n$ cells (the cells that do not carry the novel sequence); thus, the probability of this event is $(N-n)/(N+1)$. In addition, any cell that does not carry the novel sequence can become infected with the rate $\theta$.

The jump from state $n$ to state $n-1$ occurs when a cell without the novel sequence divides or a cell with the novel sequence divides and an inactivating mutations afflicts one of the daughter cells. In this case, we need to remove an infected cell from the population, an event with the probability $n/(N+1)$. Accordingly, for the birth and death rates of infected cells, we obtain:

$$\lambda_n = [(1+s)(1-u)n + \gamma N]\frac{N-n}{N+1} + \theta\frac{n(N-n)}{N}$$
$$\mu_n = [N-n + u(1+s)n]\frac{n}{N+1}$$
(4)

According to the rate equations (4), we deal with a birth and death process with a finite state space and reflecting boundaries (i.e., $\lambda_0 \neq 0, \mu_N \neq 0$). As in the deterministic model (1)-(3), we can consider the rate $\gamma$ as the rate with which uninfected individuals become infected because we can rewrite the birth rate in the following form:

$$\lambda_n = [(1+s)(1-u)n + \gamma N]\frac{N-n}{N+1} + \theta\frac{n(N-n)}{N}$$
$$\approx [(1+s)(1-u)n]\frac{N-n}{N+1} + \theta\frac{n(N-n)}{N} + \gamma(N-n)$$.

Berg and Kurland (Berg and Kurland 2002) considered a Markov process with rates (4) and $\gamma = 0$ in which there is no invasion and, accordingly, the state 0 is absorbing. Under this model, regardless of the initial distribution of the novel sequence, it will be ultimately lost in the population. In practice, the decision on whether or not to include invasion in the model rests on the comparison between the mean time to extinction from a particular initial distribution and the mean time before the appearance of a new individual carrying the novel sequence. As discussed in the last section, under conditions favoring HGT, e.g., when the donor and recipient organisms coexist in close proximity, the rate of invasion is likely to be non-negligible compared to the rate of extinction of a novel sequence. Therefore, models with invasion ($\gamma \neq 0$) are, generally, more realistic than those without it. From a formal viewpoint, inclusion of the possibility of invasion changes the qualitative behavior of the stochastic model in a way that simplifies the analysis because the stochastic model with non-zero invasion has a stable stationary distribution (a direct analog of the stable equilibrium in the deterministic model).

*2.2 Stationary distribution*

The birth and death process with a finite state space and reflecting boundaries has the unique, stable stationary distribution $p^*$ that can be easily obtained noting that the following relation must be satisfied at equilibrium:

$$\mu_n p_n^* = \lambda_{n-1} p_{n-1}^*.$$

Rearranging and iterating gives

$$p_n^* = p_k^* \prod_{j=k+1}^{n} \frac{\lambda_{j-1}}{\mu_i} = p_k^* \lambda_k \prod_{j=k+1}^{n-1} \frac{\lambda_j}{\mu_j} \frac{1}{\mu_n}. \tag{5}$$

The formula (5) is easy to evaluate numerically. If we assume that there are limits $(s+\theta)N \to h$, $\gamma N \to r$ and $uN \to q$ when $N \to \infty$, then the following theorem holds.

**Theorem 1.** *Suppose the parameters of the birth and death process with rates (4) are such that the following limits exist: $(s+\theta)N \to h$, $\gamma N \to r$, $uN \to q$ when $N \to \infty$. Then, if $N \to \infty$ and $n/N \to x$, the stationary distribution (5) asymptotically tends to the distribution with the density*

$$f(x) = C \exp(hx) x^{r-1} (1-x)^{q-1} \qquad (6)$$

*where $x \in [0,1]$ and $C$ is a constant chosen such that $\int_0^1 f(x)dx = 1$.*

The proof of this theorem is given in the Appendix.

The stationary distribution (6) is a complete analogue of the stationary distribution of the diffusion approximation of the Wright-Fisher selection-mutation model (e.g., (Goel and Richter-Dyn 1974)). The only difference is the presence of one additional parameter – the infection rate $\theta$ – which, as in the deterministic case, is added to the selection coefficient $s$.

All the qualitatively different forms of the stationary distribution can be classified using the approximation (6). This classification only depends on the products $uN$, $\gamma N$, and $(s+\theta)N$.

If $r, q < 1$ (i.e., $\gamma N$, $uN < 1$), which corresponds to a situation with low rates of inactivating mutations and invasion, the most probable states are near the boundaries. If $h > 0$ (i.e., $s + \theta > 0$), more probability is concentrated near $x = 1$ (Fig.1a), otherwise (if $h < 0$), more probability is concentrated near $x = 0$. The distribution has a single minimum. In substantive terms, if the combined values of the selective advantage and infection rate favor the survival of the novel sequence, it tends to sweep the population; otherwise, it is likely to go extinct.

If $r < 1, q > 1$ (low invasion rate, high rate of inactivating mutations), then the most probable state is near $x = 0$, i.e., the novel sequence tends to perish (Fig.1b). However, for a particular range of values $s + \theta > 0$, the distribution can have a local maximum (Fig.1c).

In the case $r > 1, q < 1$ (high invasion rate, low rate of inactivating mutations), the reverse is observed, with the most probable state being fixation of the novel sequence; however, the distribution can have a local maximum if $s + \theta < 0$ (Fig.1d,e).

Finally, if $r > 1, q > 1$ (high rates of invasion and inactivating mutations), the distribution is unimodal and this is the only case when the most probable state is close to the deterministic steady-state value for equation (3) (Fig.1f).

To summarize, the stationary distribution under the stochastic model can have at most two extremes. The most probable states are determined by the joint effect of all five parameters of the model. The approximation (6) closely mimics exact distributions produced by numerical simulation, at least for the considered ranges of parameter values (Fig.1).

Approximation (6) can help decide which parameters make substantial contributions to the evolution of the population in each particular situation. Let us consider the quantity $E\{\tilde{X}\} = g(h, r, q)$ - the average population penetration of the novel sequence (i.e., the average fraction of type 1 individuals). This is, simply, the expected value of a random variable with the density function (6). In Fig. 2a, the level lines of the function $E\{\tilde{X}\}$ are shown for a fixed $q = uN$. In Fig. 2b, the implicit function $E\{\tilde{X}\} = 0.5$ is plotted for different values of $q$. These graphs show that, if the rate of invasion $\gamma N$ is substantially lower than the rate of inactivating mutation $uN$, significant penetration (on average) can be reached only with high positive values of $(s + \theta)N$.

### 2.3. Probability of fixation

We can formally let $\lambda_0 = 0, \mu_N = 0$ to make states $0$ and $N$ absorbing. In this situation, the fate of a unique individual carrying the novel sequence can be examined. Through genetic drift, this novel sequence can be lost (the state $0$ is reached) or can penetrate all the population (the state $N$ is reached). In the latter case, we speak of fixation of the novel sequence in the population although it has to be kept in mind that, in the case of reflecting boundaries, the system does not stay in state $N$. Letting $\lambda_0 = 0, \mu_N = 0$ is somewhat artificial but the reasoning behind examining this situation is as follows. Firstly, we can assume that, once the novel sequence is acquired by all the individuals in a population, it becomes essential and cannot be lost (hence $\mu_N = 0$). When the rate of invasion is small, the typical fate of a novel sequence appearing in the population is extinction. Under typical conditions (low invasion and inactivation mutation rates), most of the time, the population waits for a "lucky" sequence to be fixed; accordingly letting $\lambda_0 = 0$ is not unrealistic. The time of fixation conditioned such that the fixation does occur is usually much lower than the mean time to reach state $N$ with a reflecting boundary at $n = 0$.

The probability that the system ends up in the state $N$ (the probability of fixation) if initially there is only one individual carrying this novel sequence is (e.g., (Goel and Richter-Dyn 1974))

$$P_{fix} = \frac{1}{1 + \sum_{i=1}^{N-1} \prod_{n=1}^{i} \frac{\mu_n}{\lambda_n}}.$$

Using approximations given in the proof of Theorem 1 (see Appendix), it can be shown that the following integral approximation can be used to evaluate $P_{fix}$:

$$P_{fix} \approx \frac{1}{1 + N^{1-\gamma N} \int_{1/N}^{1-1/N} \frac{\exp(-(s+\theta)Nx)}{x^{\gamma N}(1-x)^{uN}} dx}. \qquad (7)$$

First, let $\gamma = 0$. This case was examined by Berg and Kurland (Berg and Kurland 2002) who showed that, if $s + \theta < u$, then the probability of fixation is virtually zero for large $N$ (more precisely, this, of course, implies the effective population size $N_e$; hereinafter, we use $N$ for simplicity). If $u < s + \theta < u(1 - \ln u)$, there is a plateau where the probability of fixation does not depend on $N$, and for large $N$, there is a sharp drop in this probability. Finally, if $s + \theta > u(1 - \ln u)$, the probability of fixation has a limit that does not depend on $N$. These three distinct behaviors are illustrated in Fig. 3a.

All the curves in Fig.3a were obtained using the integral approximation (7). This integral approximation holds well in most ranges of parameters except in the region $uN \gg 1$; all analyses described in this work were well within the applicability range of the approximation. If $\gamma \neq 0$, qualitative changes in the behavior of $P_{fix}$ are observed. Obviously, $P_{fix}$ is expected to increase. However, if $\gamma < u$, i.e., the invasion rate is lower than the inactivation rate, and $s + \theta < u(1 - \ln u)$, the limiting behavior of $P_{fix}$ is the same as in the case without invasion, namely, $P_{fix} \to 0$ when $N \to \infty$ (Fig.3b,c). There is, however, a range of $N$ values in which $P_{fix}$ is greater than the same probability with $\gamma = 0$. In contrast, in the case of $s + \theta > u(1 - \ln u)$, inclusion of even low-rate invasion results in a notable change of the limiting behavior of $P_{fix}$: the plot of $P_{fix}$ versus the population size $N$ has a minimum at the point $N = N_{min}$, and for $N > N_{min}$, $P_{fix}$ increases with the increase of $N$ (Fig.3d).

The probability of fixation in the neutral case ($s = 0$) without immigration is primarily determined by the ratio of the rate of inactivating mutations, which oppose fixation, and the rate of horizontal transmission of the novel sequence within the recipient population. If the transmission (infection) rate is high enough, neutral fixation of the novel sequence is an event with a non-zero probability. The inequality $\theta > u(1 - \ln u)$ is not very restrictive because it demands that the transmission rate is approximately 10-20 times greater than the inactivation rate. Moreover, even if the novel sequence is slightly deleterious, it can be fixed not only due to random drift in a finite population but also due to the possibility of infection (Fig. 4). The fixation of slightly deleterious alleles in a finite population leading to a decline in the mean fitness of the population is known as Muller's ratchet (Muller 1964; Felsenstein 1974). The relatively high rate of the within-population transmission (infection) offers an alternative scenario for reducing the mean population fitness during evolution.

### *2.4. Quasi-stationary distributions*

If there is no invasion in the model, the novel sequence is doomed to extinct. More precisely, let $\gamma = 0$. It is readily shown that the process $\{X(t)\}$ has a degenerate stationary distribution $p^* = (1, 0, ..., 0)$. The distribution of $X(t)$ approaches the stationary distribution as time $t$ approaches infinity. Thus, ultimate absorption (extinction of the novel sequence) is certain. To evaluate the mean time to extinction if initially only one individual carries the novel sequence, we can use the following formula (e.g., (Goel and Richter-Dyn 1974)):

$$E\{T_{lost}\} = \sum_{n=1}^{N} \frac{1}{\mu_n} \prod_{j=1}^{n-1} \frac{\lambda_j}{\mu_j}.$$

Using approximations given in the proof of Theorem 1 (see Appendix), an integral approximation for this quantity can be obtained (see also (Berg and Kurland 2002) ):

$$E\{T_{lost}\} \approx \exp(-(s+\theta)) \int_{1/N}^{1} \frac{\exp(x(s+\theta)N)}{x \ (1-x)^{1-uN}} dx$$

If $s+\theta < u$, then $E\{T_{lost}\}$ is virtually the same as for the neutral expectation (when $s+\theta = 0$). In contrast, if $s+\theta > u$, i.e., selection combined with horizontal gene transmission dominates over inactivation, then the time to absorption (extinction) goes through a minimum and then increases sharply with increasing $N$ (Fig. 5) (this figure mimics Fig. 2B of Berg and Kurland (Berg and Kurland 2002)). It is therefore of interest to analyze the distribution of $X(t)$ prior to absorption. This is done using the concept of quasi-stationarity.

There are many biological and ecological systems that eventually go extinct yet appear to be stable over any reasonable time scale. The notion of quasi-stationary distribution has proved to be a powerful tool for modeling the behavior of such systems (Pollet 1996). In particular, it allows one to predict the possible distribution of $X(t)$ on its way to extinction.

The state space of the birth and death process under consideration can be partitioned into two subsets, one containing the absorbing state 0 and the other one comprised of the transient states $\{1,2,...,N\}$. Before absorption, the process assumes values in the set of transient states. If the process is conditioned on the event that absorption has not taken place at time $t$, then the conditional state probabilities $q_n(t)$ can be determined from the state probabilities $p_n(t)$ through the following relation:

$$q_n(t) = \Pr\{X(t) = n \mid X(t) > 0\} = \frac{p_n(t)}{1 - p_0(t)}.$$

By differentiating this relation and using the Kolmogorov forward equations for $p_n(t)$, we can obtain a system of differential equation for $q_n(t)$. The quasi-stationary distribution $q^*$ is the stationary solution of this system of equations. The probabilities can be shown to satisfy the following system of difference equations:

$$\mu_{n+1} q^*_{n+1} - (\lambda_n + \mu_n) q^*_n + \lambda_{n-1} q^*_{n-1} = \mu_1 q^*_1 q^*_n.$$

In the general case, these equations cannot be solved explicitly. They can, however, be used to derive the relations to which iteration methods for determining the quasi-stationary distribution can be applied (Nasell 2001) (the algorithm for evaluating the quasi-stationary distribution is presented in the Appendix). The simplest way to obtain an approximation of the quasi-stationary distribution is to restrict consideration to transient states (the state space is made strictly positive). By excluding zero from the state space, one can establish a related process without an absorbing state. This method has been applied in several mathematical models (Kendall 1949; Pielou 1969) and is valid when the time to extinction is reasonably large (Nasell 2001).

We consider an auxiliary process $\{X_0(t)\}$ which can be described as the original process with the origin removed. Formally, we put $\mu_1 = 0$, while all other transition rates are equal to the corresponding rates for the original process. The stationary distribution for the process $\{X_0(t)\}$ is easy to determine. A good approximation for this stationary distribution is given by (6) with $r = 0$ and the normalization constant determined by $\int_{1/N}^{1} f(x)dx = 1$. This means that we can use Eq. (6) as an approximation for the quasi-stationary distribution. The main question is how close this approximation to the actual quasi-stationary distribution.

Figure 6 shows that the simple approximation (6) with $\gamma=0$ is good (except for the area near 0) if $s+\theta \gg u$, i.e., this approximation holds when the mean time to extinction is sufficiently long (compare to Fig. 5). However, even with lower values of $s+\theta$, i.e, $s+\theta \sim u(1-\ln u)$, the approximation (6) is close to the observed quasi-stationary distribution almost in the whole range of x except for the area near 0 (data not shown).

Thus, even if invasion is not included in the model, the most probable states of the process can be near 1 depending on the values of the other parameters, i.e., for a certain part of the parameter space, the novel sequence, on its way to extinction, penetrates almost the entire population with a high probability.

**General discussion and conclusions**

The present model is a generalization of the stochastic model described by Berg and Kurland (Berg and Kurland 2002). The model includes five parameters: inactivating mutation rate, selection coefficient, invasion rate, within-population horizontal transmission (infection) rate, and population size. Berg and Kurland come to the conclusion that horizontally acquired sequences can be fixed in a population only when these sequences confer a substantial selective advantage onto the recipient and therefore are subject to strong positive selection. However, although they formally consider the process of infection when formulating the model, its effect is excluded from their interpretation. Therefore, their conclusions are based on the model with only two processes: genetic drift and mutational inactivation. In this setting, it becomes self-evident that, in typical, large microbial populations, horizontally transferred sequences can survive for any appreciable duration of time only when they are strongly beneficial. Since such situations are reasonably expected to be rare (the exceptions, such as acquisition of antibiotic resistance or ability to utilize new nutrients, notwithstanding), the results of Berg and Kurland's modeling imply that HGT played less of a role in the evolution of prokaryotes than it is given in "horizontal genomics" concepts (Kurland, Canback, and Berg 2003)

In the present model, two additional processes are included and, in a certain domain of parameter values, significantly contribute to the outcome. Quasi-formally, the logic can be as follows.

- If the rate of within-population horizontal transmission of the novel sequence (infection) is comparable to the inactivating mutation rate, then the mean time to extinction of this novel sequence is quite long and, importantly, increases with the increase of the population size $N$ (Fig. 5). In this case, the approximation (6) for quasi-stationary distributions is applicable. The analysis under this approximation shows that the novel sequence can penetrate a significant part of the recipient population (the most probable states are near 1 in Fig. 6).
- If the mean time to extinction is long, then the appearance of this particular novel sequence in the population may not be a unique event. Accordingly, invasion has to be taken into account, and the results obtained for the true stationary distribution are valid.
- If invasion is included in the model, then, within a reasonable time span, the novel sequence can penetrate a significant part of population (Fig. 2) and, eventually, the horizontally transferred gene may be fixed. It is interesting to note that the mean time required for significant penetration is dramatically less than the mean time of fixation (Fig. 7) which could have substantial consequences for the fate of the population.
- When a sequence persists in a population for a long time and, especially, when it gets fixed, there is a chance that the acquired gene becomes beneficial or even indispensable (essential).

The present analysis shows that taking into account the processes of within-population transmission (infection) and invasion leads to conclusions that are dramatically different from those of Berg and Kurland: if the rates of these processes are non-negligible, horizontally transferred sequences do get fixed or at least persist in a significant part of the recipient population for a long time, even if they are neutral or slightly deleterious.

The pressing question, then, is: just how likely is it that these additional processes occur at significant rates? Unfortunately, quantitative estimates are lacking which precludes us from supplementing the mathematical analysis of the model with empirical estimates as Berg and Kurland have done with regard to the inactivation rate (Berg and Kurland 2002). Qualitatively, however, biological data suggest that both processes can be mediated by several mechanisms such that their rates vary within extremely broad ranges and the gene flow could be intense under favorable conditions. Bacteria are known to exchange genes via conjugative plasmids and integrative and conjugative elements (ICEs)(Osborn and Boltner 2002; Grohmann, Muth, and Espinosa 2003; Bennett 2004; Burrus and Waldor 2004). Furthermore, it has been extensively documented that many bacteria and archaea possess the molecular machinery for non-specific DNA uptake and are highly competent for transformation which is, indeed, considered to be a major mechanism of gene transfer between microbes (Lorenz and Wackernagel 1994; Dubnau 1999; Redfield 2001; Claverys and Martin 2003; Chen and Dubnau 2004). All these processes are most intense within a microbial population, contributing to a potentially high rate of "infection" in our model. However, they are by no means limited to the same prokaryotic species and have been shown to occur even between phylogenetically distant prokaryotes (Davison 1999; Paul 1999). Such interspecies gene transfer which, again, may be intense under conditions of physical proximity between diverse prokaryotes, e.g., in microbial mats, can be a major contribution to the process of invasion in our model. That the amount of apparent HGT critically depends on ecological and physical closeness of the purported donors and recipients has been born out by comparative genomics. In particular, hyperthermophilic bacteria carry a disproportionate number of genes thought to have been horizontally acquired from archaea (Aravind et al. 1998; Nelson et al. 1999; Worning et al. 2000; Nesbo et al. 2001). Conversely, mesophilic archaea, such as *Halobacterium* and, especially, *Methanosarcina*, contain numerous genes of apparent bacterial origin, many more than hyperthermophilic archaeal species (Koonin, Makarova, and Aravind 2001; Pennisi 2001; Deppenmeier et al. 2002; Koonin et al. 2002; Koonin 2003). Numerous studies in microbial ecology reveal a remarkable diversity of microbial communities (Kassen and Rainey 2004). Microbial communities show both dynamic behavior, which is linked to niche specialization (Kerr et al. 2002), and considerable temporal stability (Fernandez et al. 1999; Fernandez et al. 2000; Hashsham et al. 2000). Generally, these communities provide fertile ground for HGT, making the models with non-zero invasion relevant.

The modeling results presented here strongly suggest that the main precept of "horizontal genomics", the crucial role of HGT in prokaryotic evolution, does not depend on the unrealistic assumption that all horizontally transferred genes that are fixed in microbial populations confer a strong selective advantage onto the recipient. We believe that this theoretical support for a major evolutionary impact of HGT is particularly important given how hard it is to obtain a rigorous proof for most HGT events. To strengthen the argument even further, it will be necessary to develop quantitative estimates for gene fluxes within and between prokaryotic populations which figure as "infection" and "invasion" in our model.

**Acknowledgements**

We thank Yuri Wolf for numerous helpful discussions and useful suggestions and Alex Kondrashov for critical reading of the manuscript.

**Figure legends**

**Fig. 1.** Possible qualitatively different stationary distributions of the HGT-mutation-selection-invasion model with rates (4). In each panel, the approximation (6) and numerically evaluated exact stationary distribution are depicted. The stationary distribution is normalized such that the areas under the curves equal one. The parameters are (a) $h = 0.06$, $r = 0.98$, $q = 0.97$; (b) $h = 0.6$, $r = 0.8$, $q = 4$; (c) $h = 6$, $r = 0.8$, $q = 4$; (d) $h = 0.06$, $r = 2$, $q = 0.8$; (e) $h = -3.8$, $r = 2$, $q = 0.8$; (f) $h = 0.06$, $r = 6$, $q = 1.8$

**Fig. 2.** (a) Contour plot of the average population penetration $E\{\tilde{X}\} = \int_0^1 xf(x)dx$ with the fixed value $uN = 0.5$. (b) Level lines for 50% population penetration of the novel sequence for different values of $uN$.

**Fig. 3**. Probability of fixation of the novel sequence as a function of the population size $N$. (a) The case of no invasion, $\gamma = 0$. The dashed line is the probability of fixation of a neutral sequence in the presence of no evolutionary forces except for genetic drift, $P_{fix} = 1/N$. The inactivating mutation rate $u = 10^{-7}$. The values of $s + \theta$ for the solid curves from bottom to top are $0, 10^{-8}, 5 \cdot 10^{-7}, 8 \cdot 10^{-7}, 5 \cdot 10^{-6}, 10^{-5}$. (b),(c),(d) The case of invasion with different parameters. (b) Parameters are $u = 10^{-7}, s + \theta = 10^{-8}$. The values of $\gamma$ for the curves from bottom to top are $0, 10^{-8}, 2 \cdot 10^{-8}, 5 \cdot 10^{-8}$; (c) Parameters are $u = 10^{-7}, s + \theta = 8 \cdot 10^{-7}$. The values of $\gamma$ for the curves from bottom to top are $0, 10^{-9}, 10^{-8}, 2 \cdot 10^{-8}$; (d) Parameters are $u = 10^{-7}, s + \theta = 5 \cdot 10^{-6}$. The values of $\gamma$ for the curves from bottom to top are $0, 10^{-10}, 10^{-9}, 5 \cdot 10^{-9}$;

**Fig. 4.** Probability of fixation of the novel sequence as a function of the selection coefficient $s$. Parameters are $N=10^8$, $u=5 \cdot 10^{-9}$, $\theta = 5 \cdot 10^{-7}$ (the upper curve), $\theta = 5 \cdot 10^{-8}$ (the lower curve). The dashed horizontal line is the probability of fixation of a neutral sequence $P_{fix} = 1/N$ when there are no evolutionary forces except for the random genetic drift

**Fig. 5.** Mean time to extinction as a function of the population size $N$. The inactivation rate $u = 10^{-7}$. The values of $s + \theta$ for the curves from bottom to top are $-10^{-8}, 10^{-8}, 10^{-7}, 5 \cdot 10^{-7}, 10^{-6}$

**Fig. 6.** Quasi-stationary distributions of the HGT-selection-mutation model. The thin solid and dotted curves are, respectively, the exact and approximate distributions of the auxiliary process with $\mu_1 = 0$, and the thick solid curves are quasi-stationary distributions obtained with iteration methods. The parameters are $(s + \theta)N = 8$, $uN = 9.5$ (a), $uN = 1.17$ (b) and $uN = 0.37$ (c).

**Fig. 7**. The mean times required to reach different levels of penetration of a novel sequence in a population. The y axis shows the ratio of the mean time to the given level of penetration $E\{T_k\}$ to the mean time to fixation $E\{T_N\}$. Parameter values: $s + \theta = 0.02$, $u = 0.0015$, $\gamma = 0.001$, $N = 3500$.

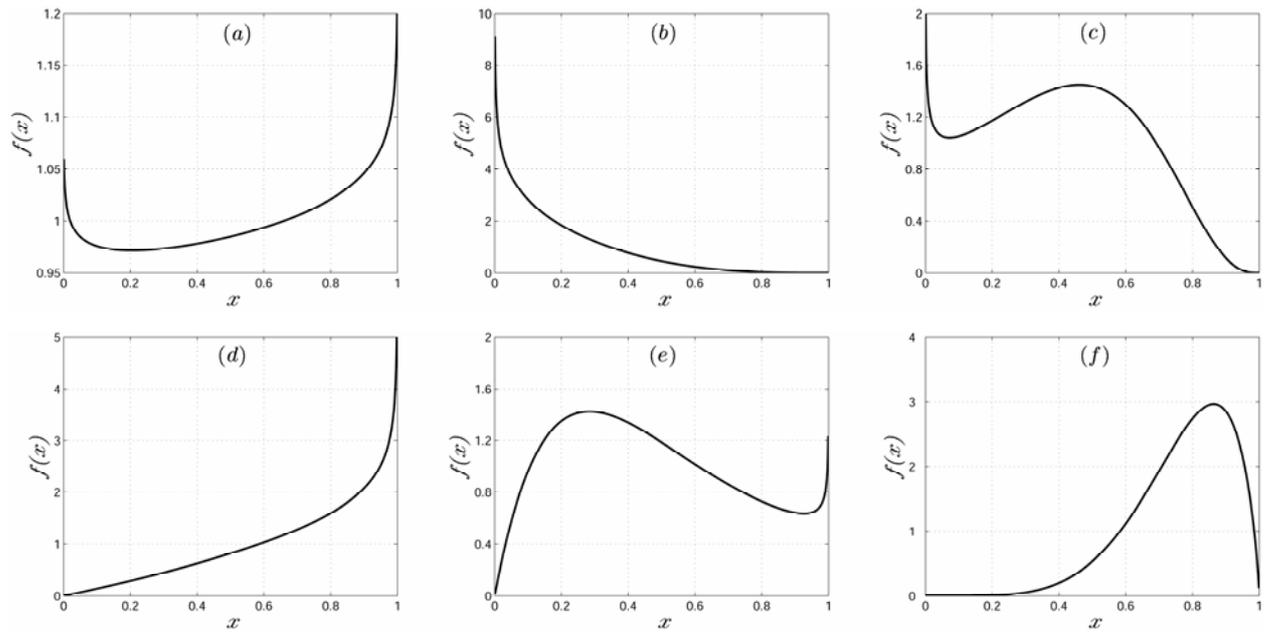

Figure 1 from Novozhilov et al., **Mathematical modeling of evolution of horizontally transferred genes**

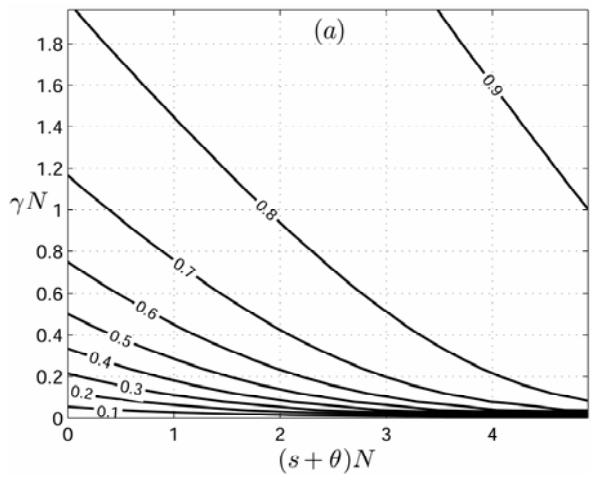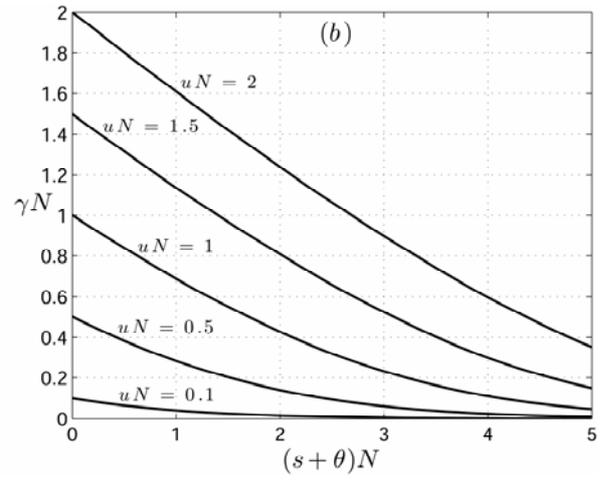

Figure 2 from Novozhilov et al., **Mathematical modeling of evolution of horizontally transferred genes**

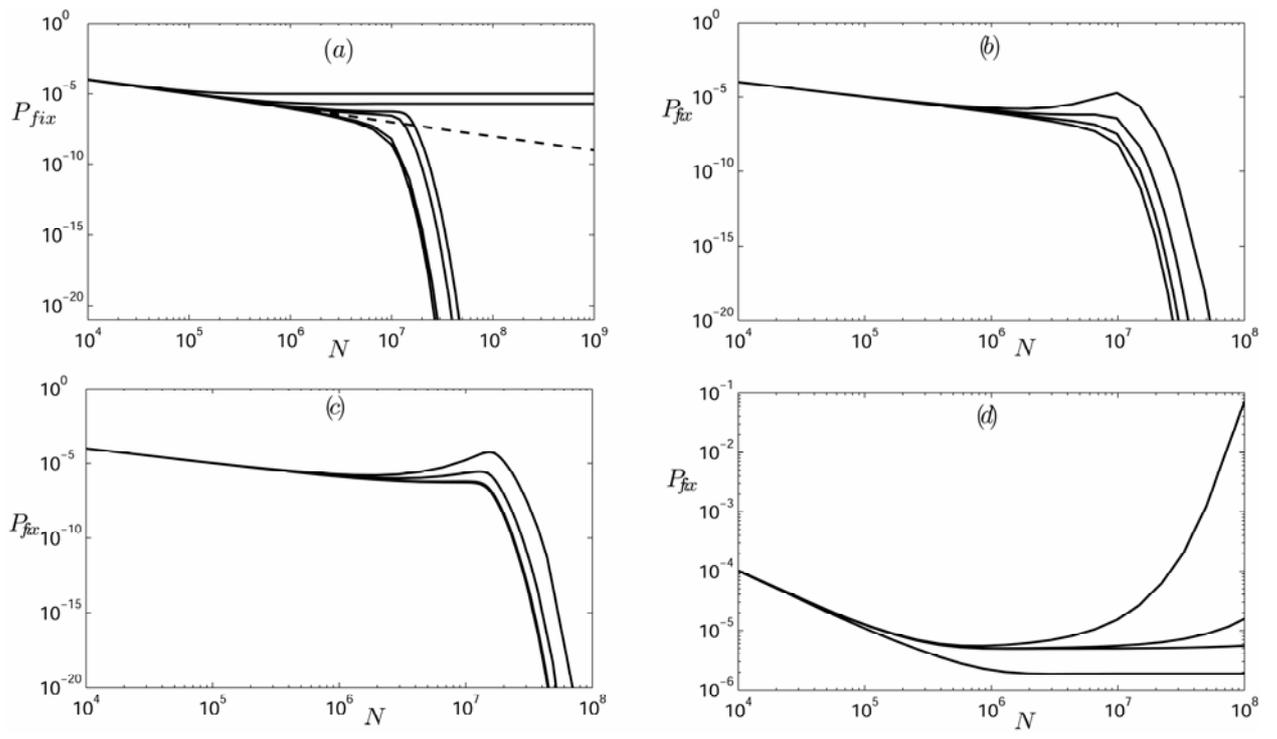

Figure 3 from Novozhilov et al., **Mathematical modeling of evolution of horizontally transferred genes**

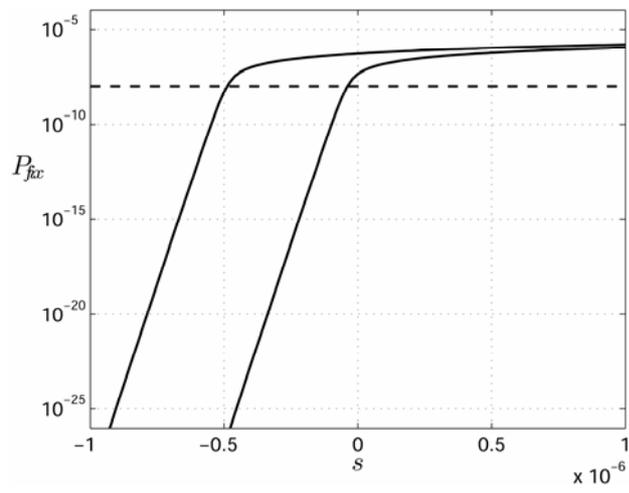

Figure 4 from Novozhilov et al., **Mathematical modeling of evolution of horizontally transferred genes**

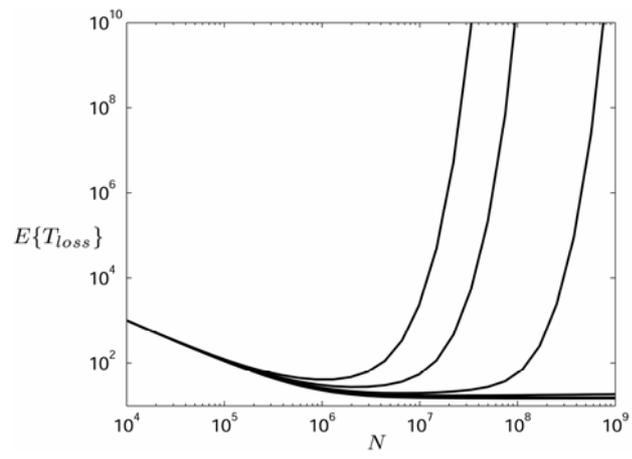

Figure 5 from Novozhilov et al., **Mathematical modeling of evolution of horizontally transferred genes**

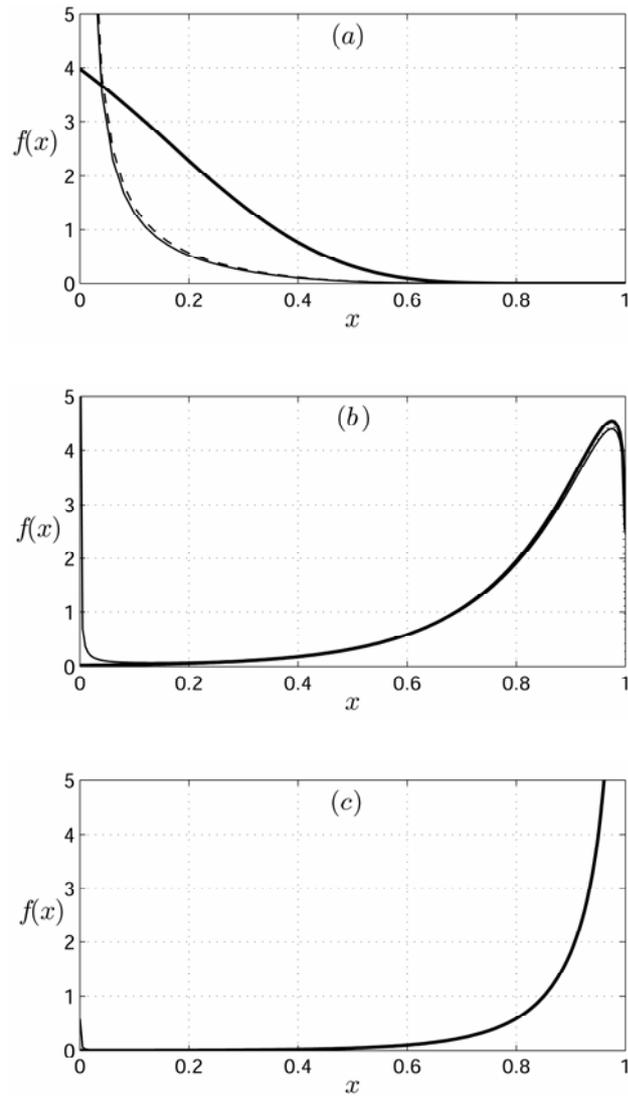

Figure 6 from Novozhilov et al., **Mathematical modeling of evolution of horizontally transferred genes**

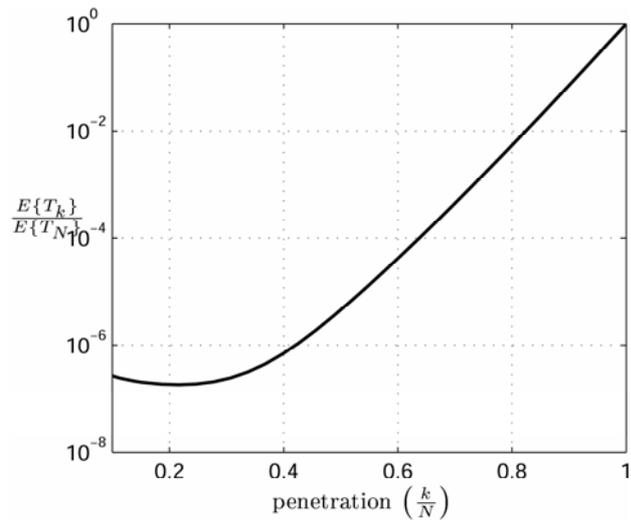

Figure 7 from Novozhilov et al., **Mathematical modeling of evolution of horizontally transferred genes**

**Appendix**

# Mathematical modeling of evolution of horizontally transferred genes

Artem S. Novozhilov, Georgy P. Karev, and Eugene V. Koonin*

National Center for Biotechnology Information, National Library of Medicine, National Institutes of Health, Bethesda, MD 20894

## 1. Proof of Theorem 1

First note that, if $n/N \to x$ when $N \to \infty$, then

$$\mu_n \to Nx(1-x). \tag{a}$$

To handle the product in (5), we note that

$$\begin{aligned}
\frac{\lambda_j}{\mu_j} &= \frac{(N-j)\left[(1+s)(1-u)j + \gamma N + \theta j \frac{N+1}{N}\right]}{j[N-j+u(1+s)j]} \\
&\approx \frac{(N-j)((1+s-u+\theta)j + \gamma N)}{j(N-j+uj)} \\
&= \frac{1+s-u+\theta+\frac{\gamma N}{j}}{1-u+\frac{uN}{N-j}} \approx 1+s+\theta+\frac{\gamma N}{j}-\frac{uN}{N-j}
\end{aligned}$$

where we used the fact that $(1+x)/(1-y) \approx 1+x+y$ for small $x$ and $y$. Taking the logarithm of (5), we obtain

$$\begin{aligned}
\ln\left(\frac{p_n^* \mu_n}{p_k^* \lambda_k}\right) &\approx \sum_{j=k+1}^{n-1} \ln\left(1+s+\theta+\frac{\gamma N}{j}-\frac{uN}{N-j}\right) \\
&\approx \sum_{j=k+1}^{n-1}\left[s+\theta+\frac{\gamma N}{j}-\frac{uN}{N-j}\right] \\
&\approx (s+\theta)(n-k-1) \sum_{j=k+1}^{n-1}\left[\frac{r}{j/N}-\frac{q}{1-j/N}\right]\frac{1}{N} \\
&\approx (s+\theta)(n-k-1) \int_{k/N}^{n/N}\left[\frac{r}{x}-\frac{q}{1-x}\right]dx \\
&\approx (s+\theta)(n-k-1)\ln\left(x^r(1-x)^q\right)\Big|_{x=k/N}^{x=n/N}
\end{aligned}$$

Here we used approximation $\ln(1+x) \approx x$ for small $x$.

Taking $k = N/2$ as a reference point and noting that $\lambda_{N/2} \to N/4$ as $N \to \infty$ we obtain

$$p_n^* \approx \frac{N p_{N/2}^*}{\mu_n} 2^{r+q+2} e^{h(n/N-3/2)} (n/N)^r (1-n/N)^q$$

Using (a) we obtain the desired result. This completes the proof.

## 2. Algorithm for calculating a quasi-stationary distribution

Here we follow (Nasell 2001). We examine a birth-death process $\{X(t), t \geq 0\}$ with the finite state space $\{0, 1, ..., N\}$ where the origin is an absorbing state. The birth rate is denoted $\lambda_n$ and the death rate is denoted $\mu_n$. The Kolmogorov forward equations for the state probabilities $p_n(t) = \Pr\{X(t) = n\}$ can be written as

$$\dot{p}_n = \mu_{n+1} p_{n+1} - (\lambda_n + \mu_n) p_n + \lambda_{n-1} p_{n-1}, \quad n = 0, 1, ..., N.$$

To interpret this equation, we formally put $\lambda_{-1} = \mu_{N+1} = p_{-1}(t) = p_{N+1}(t) = 0$. The state probabilities depend on the initial distribution $p_n(0)$.

Let us introduce two sequences $\rho_n$ and $\pi_n$ as follows:

$$\rho_1 = 1, \quad \rho_n = \frac{\lambda_1 \lambda_2 ... \lambda_{n-1}}{\mu_1 \mu_2 ... \mu_{n-1}}, \quad n = 2, 3, ..., N,$$

$$\pi_n = \frac{\mu_1}{\mu_n} \rho_n, \quad n = 1, 2, ..., N.$$

The state space of the birth and death process under consideration can be partitioned into the union of two subsets, one containing the absorbing state zero, and the other equal to the set of transient states $\{1, 2, ..., N\}$. Before absorption, the process assumes values in the set of transient states. If the process is conditioned on the event that absorption has not taken place at time $t$, then the conditional state probabilities $q_n(t)$ can be determined from the state probabilities $p_n(t)$ through the following relation:

$$q_n(t) = \Pr\{X(t) = n \mid X(t) > 0\} = \frac{p_n(t)}{1 - p_0(t)}.$$

By differentiating this relation and using of the Kolmogorov forward equations for $p_n(t)$, we can obtain a system of differential equation for $q_n(t)$. The quasi-stationary distribution $q^*$ is the stationary solution of this system of equations. The probabilities can be shown to satisfy the following system of difference equations:

$$\mu_{n+1}q^*_{n+1} - (\lambda_n + \mu_n)q^*_n + \lambda_{n-1}q^*_{n-1} = \mu_1 q^*_1 q^*_n.$$

It can be shown that

$$q_n = \pi_n \sum_{k=1}^{n} \frac{1 - \sum_{j=1}^{k-1} q_j}{\rho_k} q_1, \quad n = 1, 2, \ldots, N, \quad \sum_{n=1}^{N} q_n = 1.$$

This is not an explicit solution because $q_1$ can only be determined when all $q_n$ are known. However, this relation can be used in an iterative algorithm in order to obtain a quasi-stationary distribution. This algorithm starts with an arbitrary initial quasi-stationary distribution, employs this distribution as an input in the numerators of the terms that are summed over $k$ and solves the equation under the requirement that $\sum_{n=1}^{N} q_n = 1$. The iteration can be formally described as follows:

$$q_n^{(i+1)} = \pi_n \sum_{k=1}^{n} \frac{1 - \sum_{j=1}^{k-1} q_j^i}{\rho_k} q_1^{(i+1)},$$

where the superscript $i$ denotes the iteration number. The process is repeated until the results of successive iterations are sufficiently close.